\documentclass[a4paper]{article}
\usepackage[dvips]{graphicx}
\usepackage{amssymb}

\begin{document}

\title{Phase diagram of the lattice \(SU(2)\) Higgs model}
\author{C.~Bonati, G.~Cossu, M.~D'Elia, A.~Di Giacomo}
\date{}
\maketitle

\begin{abstract}
We perform a detailed study of the phase diagram of the lattice Higgs \(SU(2)\) model with 
fixed Higgs field length. Consistently with previsions based on the Fradkin-Shenker theorem we 
find a first order transition line with an endpoint whose position we determined. The diagram also 
shows cross-over lines: the cross-over corresponding to the pure \(SU(2)\) bulk is also present
at nonzero coupling with the Higgs field and merges with the one that continues the line of first
order transition beyond the critical endpoint. At high temperature the first order line becomes
a crossover, whose position moves by varying the temperature.
\end{abstract}

\section{Introduction}

In this work we present a detailed study of the phase diagram of the lattice Higgs 
\(SU(2)\) model with Higgs field in the fundamental representation and of
fixed length\footnote{A first report of this study was presented at the LATTICE 2008 
conference, \cite{proceed}.}. The model in which
Higgs length is allowed to change has received quite a lot of attention in the past for 
its possible phenomenological implications (see e.g. \cite{Fodor}), so that its phase 
diagram is known with good precision. No systematic study exists instead for the case in which the 
Higgs length is frozen. However this last model is often used, mainly because of its 
computational simplicity in numerical simulations, as the prototype of a non-Abelian gauge theory 
coupled with matter in the fundamental representation. In particular some work has been done to 
study confinement in this model (\cite{GreensiteOlejnik} and \cite{CaudyGreensite}). In those 
works some properties of the phase diagram are usually taken for granted, like the existence 
of a line of first order transitions for  \(\beta\gtrsim 2.3\), but they have never really been
tested in simulations. As a first step towards a complete understanding of this model, we thus 
started to systematically investigate its phase diagram, in order to obtain precise 
estimates of the location of its critical points.

The action of lattice Higgs \(SU(2)\) model we adopt is 
\begin{equation}\label{eq1}
S=\beta\sum_{x, \mu<\nu}\left\{1-\frac{1}{2}\mathrm{Re}\mathrm{Tr} P_{\mu\nu}(x)\right\}-
\frac{\kappa}{2}\sum_{x, \mu>0}\mathrm{Tr}[\phi^{\dag}(x)U_{\mu}(x+\hat{\mu})\phi(x+\hat{\mu})]
\end{equation}
where the first term is the standard Wilson action and the Higgs field \(\phi\) (which transforms 
in the fundamental representation) is rewritten as an \(SU(2)\) matrix (see e.g. \cite{Montvay}, 
\cite{MontvayMunster}). Since the action is linear in the fields at each point, standard 
heatbath (\cite{Creutz80}, \cite{KennedyPendleton}) and overrelaxation (\cite{Creutz87}) 
algorithms can be used for the Monte Carlo update. If the Higgs length is allowed to change this 
is no longer true, since in this case a quartic term is needed, which has the form 
\(\lambda\{\frac{1}{2}\mathrm{Tr}[\phi^{\dag}(x)\phi(x)] -1\}^2\) and destroys linearity.

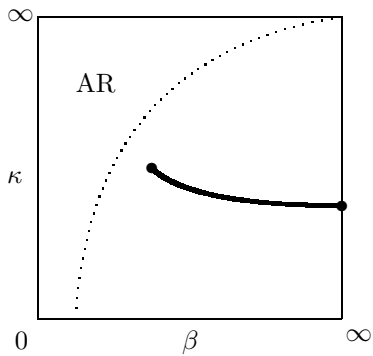
\begin{figure}[ht]
\begin{center}
\setlength{\unitlength}{1cm}
\begin{picture}(4, 4)
 \put(0, 0){\line(0, 1){4}}
 \put(0, 0){\line(1, 0){4}}
 \put(0, 4){\line(1, 0){4}}
 \put(4, 0){\line(0, 1){4}}
 \put(4, 1.5){\circle*{.15}}
 \put(1.5, 2){\circle*{.15}}
 \put(1.9,-.4){\(\beta\)}
 \put(-.3,-.4){\(0\)}
 \put(4.05,-.3){\(\infty\)}
 \put(-.4,1.8){\(\kappa\)}
 \put(-.4,3.95){\(\infty\)}
 \put(0.5,3){AR}
 \qbezier[50](.5, 0)(.7, 3.5)(4, 4)
 \linethickness{0.5mm}
 \qbezier(1.5, 2)(2, 1.5)(4, 1.5)
\end{picture}
\end{center}
\caption{Phase diagram of the Higgs \(SU(2)\) model as predicted in \cite{FradkinShenker}.
\label{fig_phase}}
\end{figure}

In the limit \(\beta\to\infty\) the theory reduces to an \(O(4)\) non-linear sigma model, which 
is known to have a mean field phase transition for \(\kappa\approx 0.6\) (see e.g. 
\cite{HasenfratzEtAl}); in \cite{FradkinShenker} it was shown that this second order transition 
becomes a first order  \emph{\`a la} Coleman-Weinberg  when the gauge field is introduced as a 
small perturbation (\(\beta\) large). The authors of \cite{FradkinShenker}, using the cluster 
expansion developed  in \cite{OsterwalderSeiler}, were also able to prove the existence of a 
region of parameter space near the axis \(\beta=0\) and for \(\kappa\to\infty\) where 
\emph{every local observable is analytic}. This statement is often referred to as the Fradkin-Shenker 
(FS) theorem. It is important to notice that, in this context, an observable is defined 
as local when its support is contained in a compact set in the thermodynamic limit; observables
not satisfying this requirement can have non-analytic behaviour (see e.g. \cite{Bertle, Grady}). 
This two results suggest a phase diagram like that shown in Fig. \ref{fig_phase}: 
the analyticity region is indicated by AR and is limited by the dotted line, the thick line is 
the line of first order transitions and the two dots are its second order endpoints. As long as 
we consider the model at zero temperature, the \(\epsilon\)-expansion predicts the endpoints to 
be in the mean field universality class.

A similar phase diagram was observed in the works \cite{Fodor}, \cite{Bock} for the lattice Higgs 
\(SU(2)\) model with fourth-order scalar coupling \(\lambda\le 0.5\), while in the model 
considered here \(\lambda\to\infty\). Because of the supposed triviality of the \(\phi^4\) model 
in four dimensions, \(\lambda\) is expected to be a marginally irrelevant parameter and therefore 
the phase diagram not to change qualitatively for \(\lambda\to\infty\) (see e.g. 
\cite{Callaway}); however it was observed since long time that the first order transition gets 
weaker as \(\lambda\) is increased, so that the phase diagram of Fig. \ref{fig_phase} has not 
really been checked at large values of \(\lambda\). 

After the seminal work ref. \cite{LangRebbiVirasoro} on very small lattices, in ref. \cite{LangguthMontvay} 
the observation of a double peak structure was reported at \(\beta=2.3\) on a \(16^4\) 
lattice, which however was probably only a consequence of the poor statistics, since in a 
later study, ref. \cite{Campos}, no double peak was found at \(\beta=2.3\). There it was stated that
``the system exhibits a transient behaviour up to \(L=24\) along which the order of the 
transition cannot be discerned''. In this paper we produce the first clear evidence of the 
line of first order transition and we obtain an estimate of the endpoint position.

\section{Numerical results}

The obvious observables to look at for this system are
\begin{itemize}
\item the gauge-Higgs coupling, \(G=\frac{1}{2}\langle\mathrm{Tr}[\phi^{\dag}(x)
U_{\mu}(x+\hat{\mu})\phi(x+\hat{\mu})]\rangle\)
\item the plaquette, \(P=\frac{1}{2}\langle\mathrm{Tr}P_{\mu\nu}\rangle\)
\item the energy density \(E=6\beta P+4\kappa G\)
\end{itemize}
Besides these natural ones, we monitored also the following observables:
\begin{itemize}
\item the Polyakov loop \(P_L(\vec{x})=\frac{1}{2}\mathrm{Tr}\left[\prod_{t=0}^{L_t-1}
      U_0(t, \vec{x})\right]\), \(P_L=\frac{1}{V}\langle\sum_{\vec{x}}P_L(\vec{x})\rangle\)
\item the \(Z_2\) monopoles density, \(M=1-\frac{1}{N_c}\sum_c \sigma_c\), where \(c\) stand for 
      the elementary cube and \(\sigma_c=\prod\limits_{P_{\mu\nu}\in\partial c}\mathrm{sign}\,
      \mathrm{Tr}P_{\mu\nu}\)
\end{itemize}
The Polyakov loop behaviour is used as an indicator of confinement. The study of the \(Z_2\) 
monopoles is motivated by the similarity of the first order transition with the bulk transition 
of the  \(SU(2)\) pure gauge theory, which is driven by lattice artefacts such as the \(Z_2\) 
monopoles. Both these points will be discussed more accurately in the following.

Data were analyzed by using the optimized histogram method (\cite{FerrenbergSwendsen}) and the 
statistical errors were estimated by using the moving block bootstrap method (see e.g. 
\cite{MignaniRosa}).

The presentation of the simulation results will be divided in several parts
\begin{enumerate}
\item we will show that at \(\beta=2.5\) there is no signal of a phase transition and data are
consistent with a smooth cross-over
\item we will show that for \(\beta\ge 2.775\) the scaling is consistent with a 
first order transition and we will present evidence of a double peak structure
\item two independent estimates of the endpoint will be obtained
\item we will give hints that the above transitions are not related to confinement
\item we will investigate the relation between the line of first order transition, which becomes 
a smooth cross-over beyond the endpoint, and the pure \(SU(2)\) bulk transition 
\item finally we will present some exploratory results at \(T\not=0\)
\end{enumerate}

All results in the first four parts have been obtained by fixing the value of \(\beta\) 
and by looking for transitions in \(\kappa\).

\subsection{Cross-over region}

\begin{figure}[p]
\begin{center}
\scalebox{0.3}{\rotatebox{-90}{\includegraphics{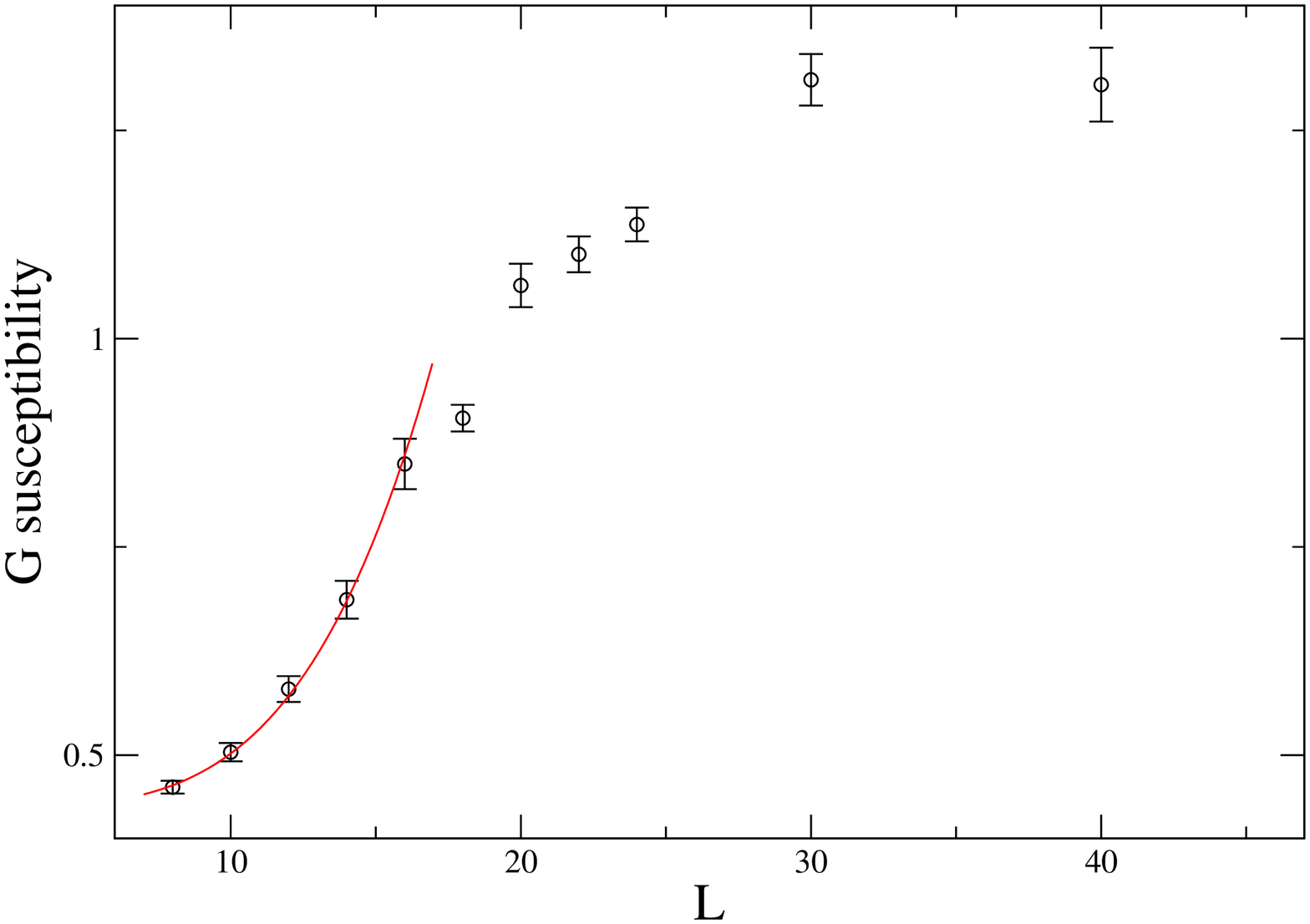}}}
\scalebox{0.3}{\rotatebox{-90}{\includegraphics{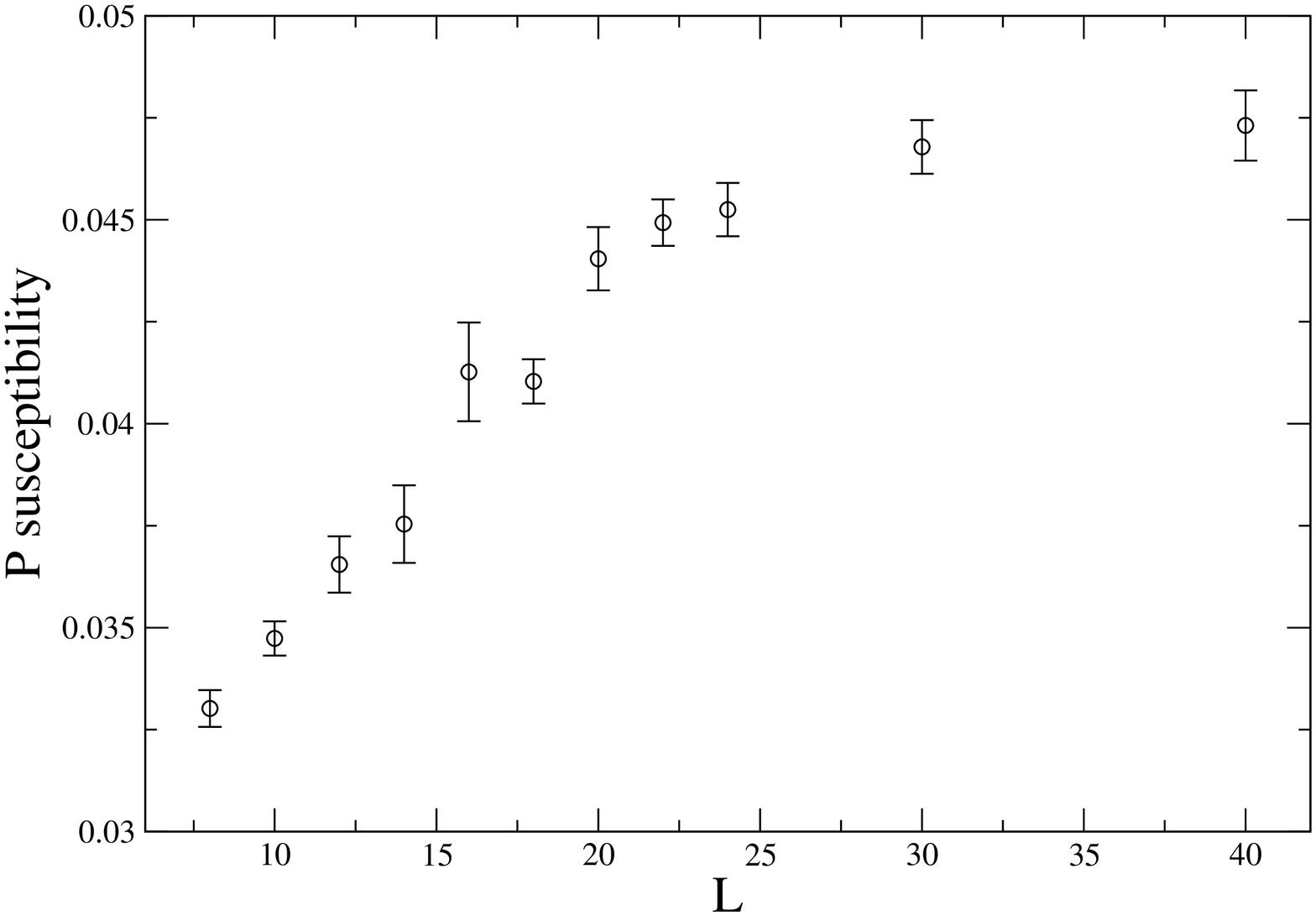}}}
\scalebox{0.3}{\rotatebox{-90}{\includegraphics{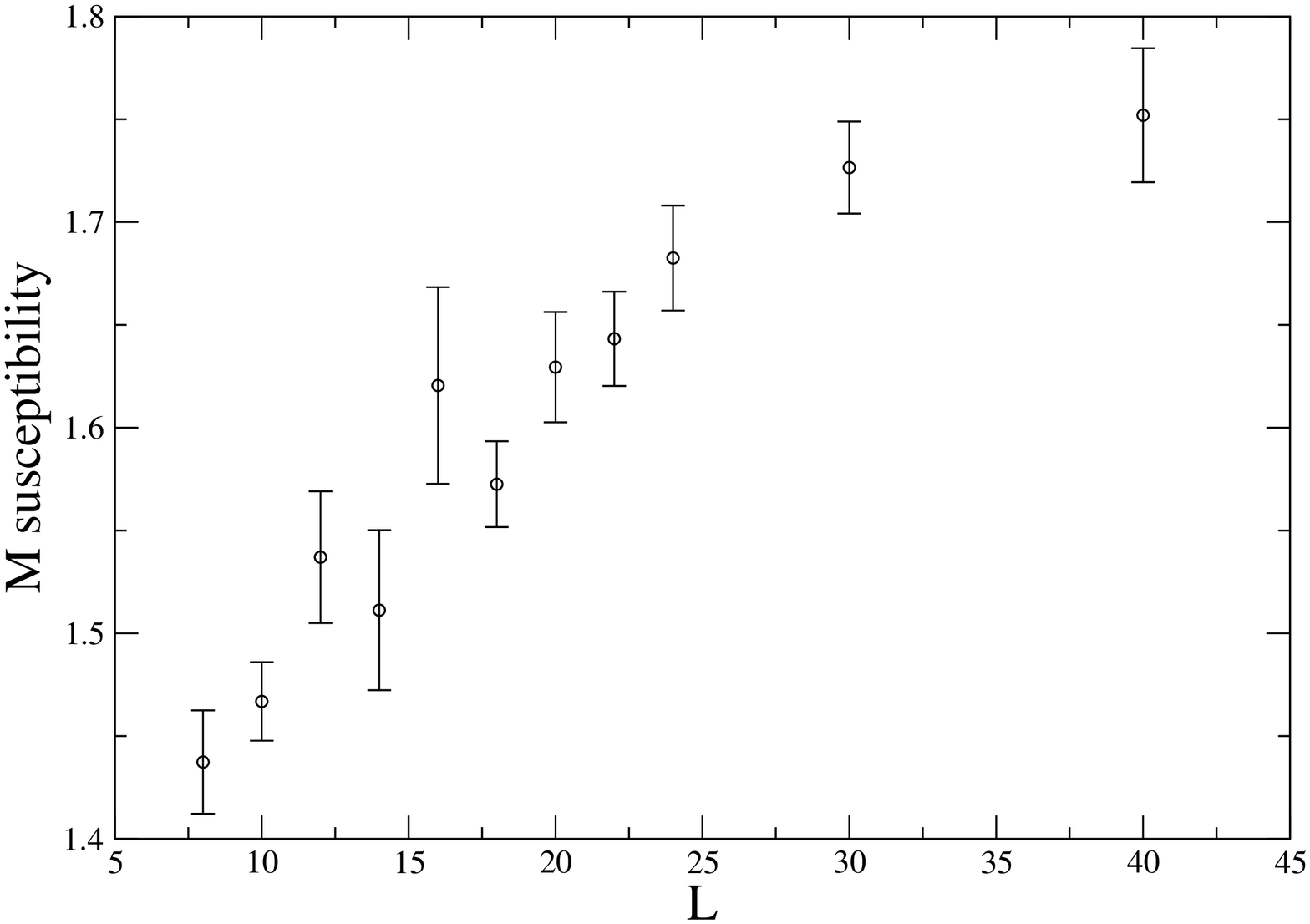}}}
\caption{\emph{Top}: \(G\) susceptibility peak heights at \(\beta=2.5\) (the continuous line is 
a fit to \(a+bx^4\)). \emph{Center}:  \(P\) susceptibility peak heights at \(\beta=2.5\).  
\emph{Bottom}: \(M\) susceptibility peak heights at \(\beta=2.5\). Clearly they all saturate at 
large \(L\).\label{fig_max_2.5}}
\end{center}
\end{figure}

In Fig. \ref{fig_max_2.5} the maxima of the susceptibilities of \(G\), \(P\) and \(M\) are 
plotted for various lattice sizes and \(\beta=2.5\). For lattices up to \(L\approx 18\) 
the maxima of susceptibilities are well described by a function of the form \(a+bL^4\), so that 
they seem to scale linearly with volume as expected for a first order transition. 
However on larger lattices all susceptibilities saturate and no singularity seems to develop 
in the thermodynamical limit. This means that the system has a correlation length of order 
\(\approx 10\) lattice spacing, so that the increase of the susceptibilities with volume, that 
in previous studies was interpreted as due to a first order transition, is just a signal that 
the lattices were too small.

To look for transitions, besides susceptibilities we also check the behaviour of the energy 
Binder cumulant, which is defined as \(B=1-\langle E^4\rangle/(3\langle E^2\rangle ^2)\). It 
can be shown (see e.g. \cite{LeeKosterlitz}) that near a transition \(B\) develops minima 
whose depth scales as
\begin{equation}\label{eq2}
B|_{min}=\frac{2}{3}-\frac{1}{12}\left(\frac{E_+}{E_-}-\frac{E_-}{E_+}\right)^2+O(L^{-4})=
\frac{2}{3}-\frac{1}{3}\left(\frac{\Delta}{\epsilon}\right)^2+O(\Delta^3)+O(L^{-4})
\end{equation}
where \(E_{\pm}=\lim_{\beta\to\beta_c^{\pm}}\langle E\rangle\), \(\Delta=E_+-E_-\) and 
\(\epsilon=\frac{1}{2}(E_++E_-)\). In particular the thermodynamical limit of  \(B|_{min}\) is less 
than \(2/3\) if and only if a latent heat is present.

\begin{figure}[h]
\begin{center}
\scalebox{0.35}{\rotatebox{-90}{\includegraphics{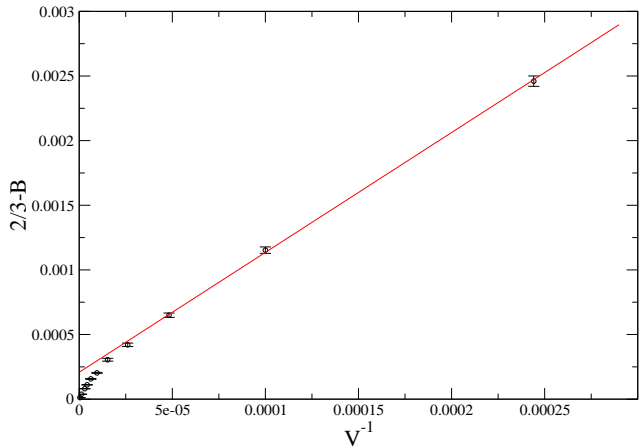}}}
\caption{Binder cumulant minima of \(E\) at \(\beta=2.5\) (\(V=L^4\)). \label{fig_binder2.5}}
\end{center}
\end{figure}

The scaling of the energy Binder cumulant minima at \(\beta=2.5\)  is shown in Fig. 
\ref{fig_binder2.5}. Also here two different behaviours are clearly visible: on small lattices 
there is scaling consistent with a first order transition. By increasing the volume \(B\to 2/3\),  
indicating a smooth cross-over or a second order transition.

These results indicate that at \(\beta=2.5\) there is no transition; this is in sharp contrast 
with all previous studies of this model, which concluded that for \(\beta \ge 2.3\) the system 
undergoes a first order transition. This wrong conclusion was based on the analysis of 
lattices of size up to \(25^4\), which we have just shown to be too small.

\begin{figure}[p]
\begin{center}
\scalebox{0.4}{\rotatebox{0}{\includegraphics{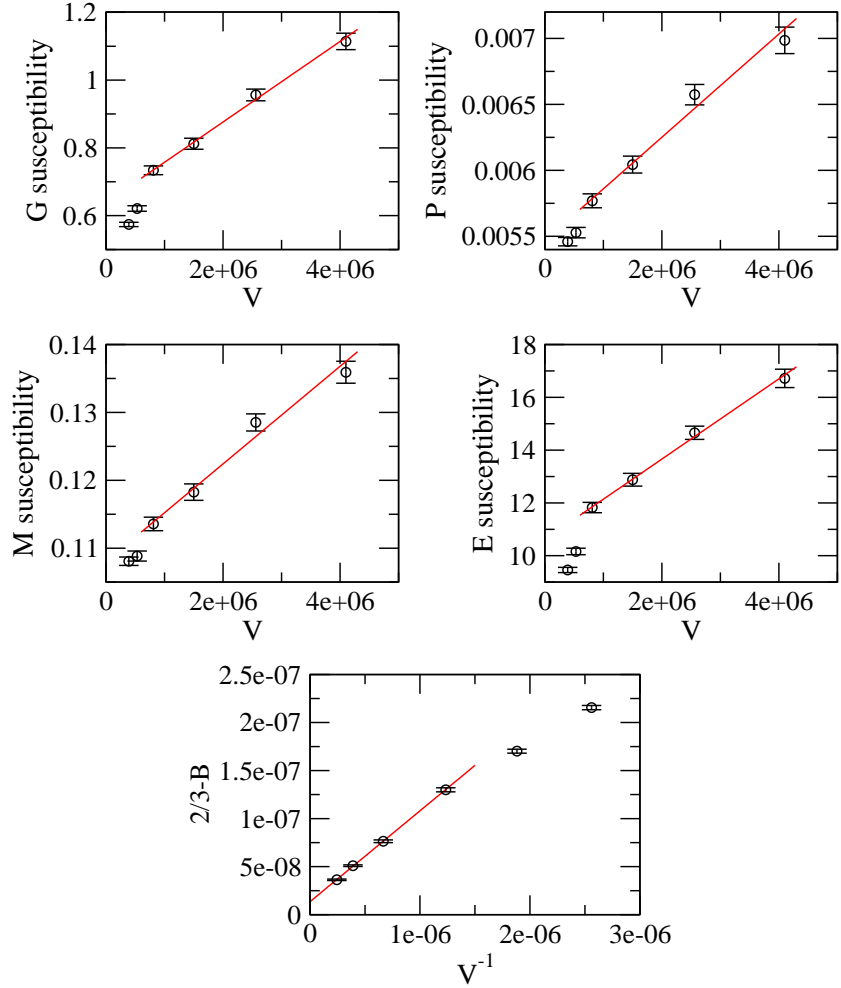}}}
\caption{Scaling of the maxima of the susceptibilities and of the \(E\) Binder cumulant minima 
for \(\beta=2.775\) (\(V=L^4\) and the larger \(V\) values correspond to lattice sizes 
\(L=30, 35, 40, 45\)). \label{fig_max2.775}}
\end{center}
\end{figure}

\subsection{First order region}

At \(\beta=2.775\), \(\beta=2.79\) and \(\beta=2.8\) the scaling of the susceptibilities and 
of the energy Binder cumulant remains consistent with first order also for the larger lattices, 
as shown for example in Fig. \ref{fig_max2.775} for \(\beta=2.775\). In this range of \(\beta\) 
values the transition gets stronger as \(\beta\) increases. However we know that the transition 
at \(\beta\to\infty\) has to become of second order, so that going at \(\beta\) high enough the 
transition has to get weaker. We could not reach this regime in our simulations since when
increasing the \(\beta\) value it is also necessary to use larger lattices. To have results free
of spurious finite size transitions the lattice must be large enough for the corresponding pure 
\(SU(2)\) gauge theory to be confined and exponentially large lattices in \(\beta\) are needed.

\begin{figure}[t]
\begin{center}
\scalebox{0.4}{\rotatebox{-90}{\includegraphics{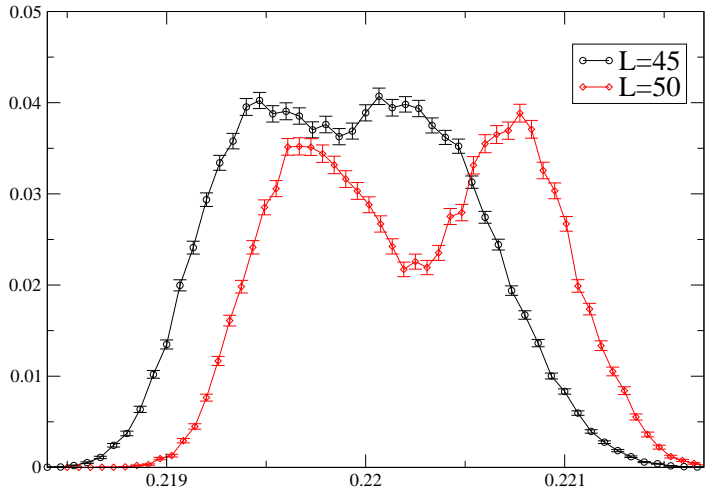}}}
\caption{Histogram of the observable \(G\) for \(\beta=2.8\) on the two largest lattices.
\label{metastability}}
\end{center}
\end{figure}

Although all observables scale consistently with a first order transition, a clear signal of 
metastability was revealed only at \(\beta=2.8\), where the transition is stronger, and only in 
the two largest lattices, namely \(L=45\) and \(L=50\); the histograms for the observable \(G\) 
on these two lattices are shown in Fig. \ref{metastability}, where the formation of a double 
peak structure is visible.

\subsection{Endpoint}

\begin{figure}[p]
\begin{center}
\scalebox{0.45}{\rotatebox{-90}{\includegraphics{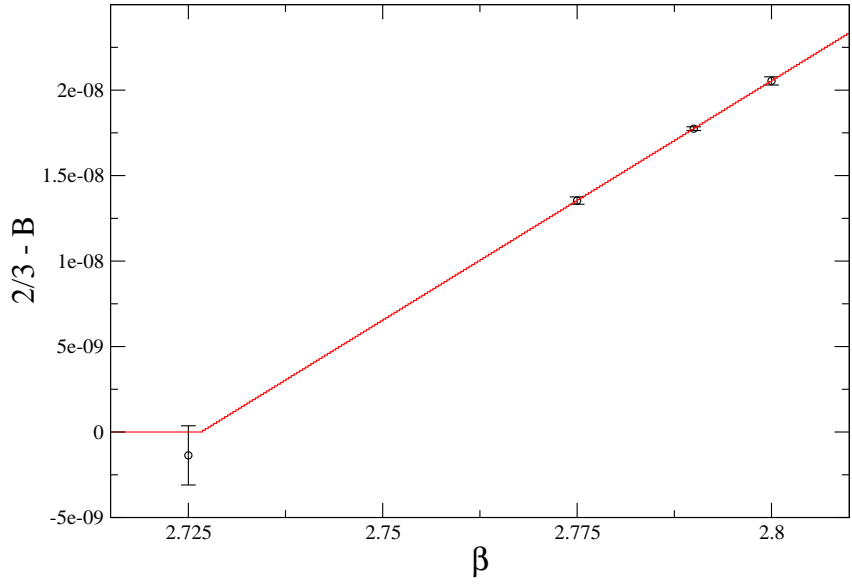}}}
\scalebox{0.45}{\rotatebox{-90}{\includegraphics{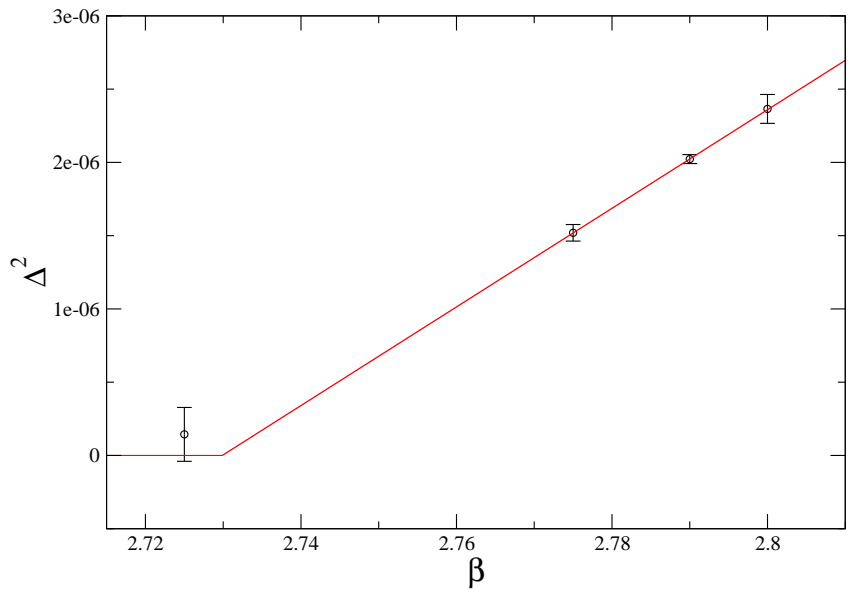}}}
\caption{\emph{Top}: Determination of the critical point using the method (1). \emph{Bottom}: 
Determination of the critical point using the method (2). 
\label{endpoints}}
\end{center}
\end{figure}

Having three sets of data with first order scaling and knowing the universality class 
of the endpoint, we can estimate its position \(\beta_c\) assuming that we are close enough to it.
We can do this in two independent ways: 
\begin{enumerate}
\item From the second form of equation \ref{eq2} we know the dependence of \(B\) on the 
discontinuity \(\Delta\) and for a mean-field transition we have \(\Delta\propto t^{1/2}\), 
where \(t\) is the reduced temperature, \(t=(T-T_c)/T_c\). In our case, near the critical point, we can use 
\(t\propto (\beta-\beta_c)\) and therefore equation \ref{eq2} can be written 
as\footnote{A subtlety has to be considered here: 
since in this model the two relevant operators at the mean field critical point are not related 
to any symmetry of the system, there is in general mixing between the magnetic and thermal 
relevant operators (see e.g. \cite{Wilding}). However, since the \(\Delta\)'s are 
measured along the coexistence curve, near the critical endpoint the mixing is negligible and 
the relation between \(\Delta\) and \(t\) is as usual \(\Delta\propto t^{\beta}\) 
(here ``\(\beta\)'' is the \(\beta\) critical exponent).}
\begin{equation}
\lim_{L\to\infty}B_L|_{min}=\frac{2}{3}-\gamma_1 (\beta-\beta_c)+O(\beta-\beta_c)^2
\end{equation}
\item For a first order transition the susceptibilities scale for large volume as         
\begin{equation}
\chi=V(\langle O^2\rangle -\langle O\rangle^2)\approx \mathrm{const}+ V\Delta^2
\end{equation}
so that fitting the \(\chi\) maxima to the relation \(\chi=\mathrm{const}+V\Delta^2\) and using 
again \(\Delta\propto t^{1/2}\), we obtain 
\begin{equation}
\Delta^2=\gamma_2(\beta-\beta_c)+O(\beta-\beta_c)^2
\end{equation}
\end{enumerate}

The fit in the planes \((\beta, B)\) and \((\beta,\Delta^2)\) are shown in Fig. \ref{endpoints}
(we use the \(G\) susceptibility); in both cases the fit is very good  and the estimates 
for the critical point position are
\begin{equation}
\beta_c^{(1)}=2.7266(16)\qquad\beta_c^{(2)}=2.7299(36)
\end{equation}
which agree with each other within errors.

\subsection{Relation to confinement}

The next question is  to understand if the two ``phases'' found at large \(\beta\) have 
different confinement properties. We recall that they cannot be considered as different 
thermodynamical phases as a consequence of FS theorem. 

\begin{figure}[p]
\begin{center}
\scalebox{0.29}{\rotatebox{-90}{\includegraphics{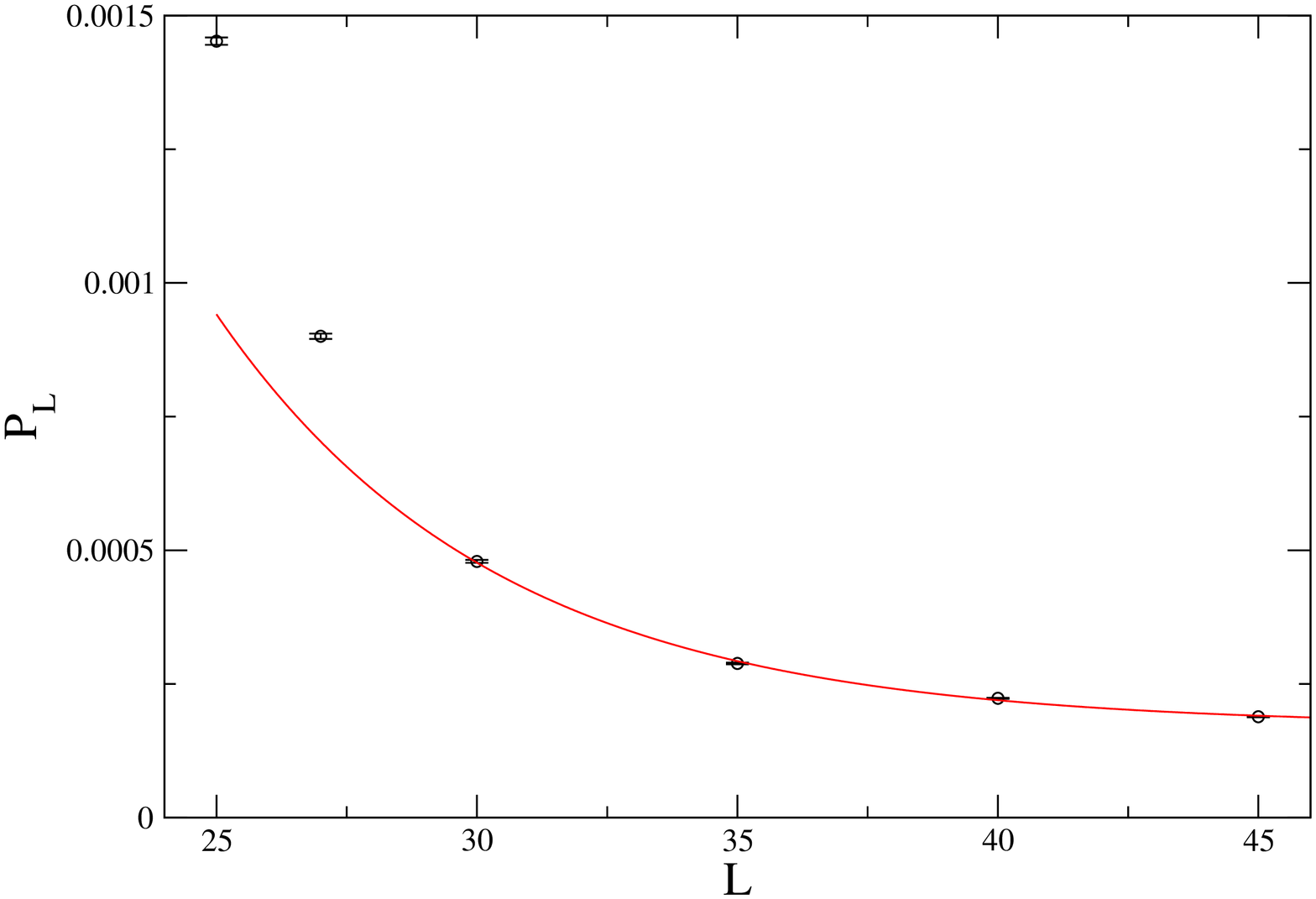}}}
\scalebox{0.29}{\rotatebox{-90}{\includegraphics{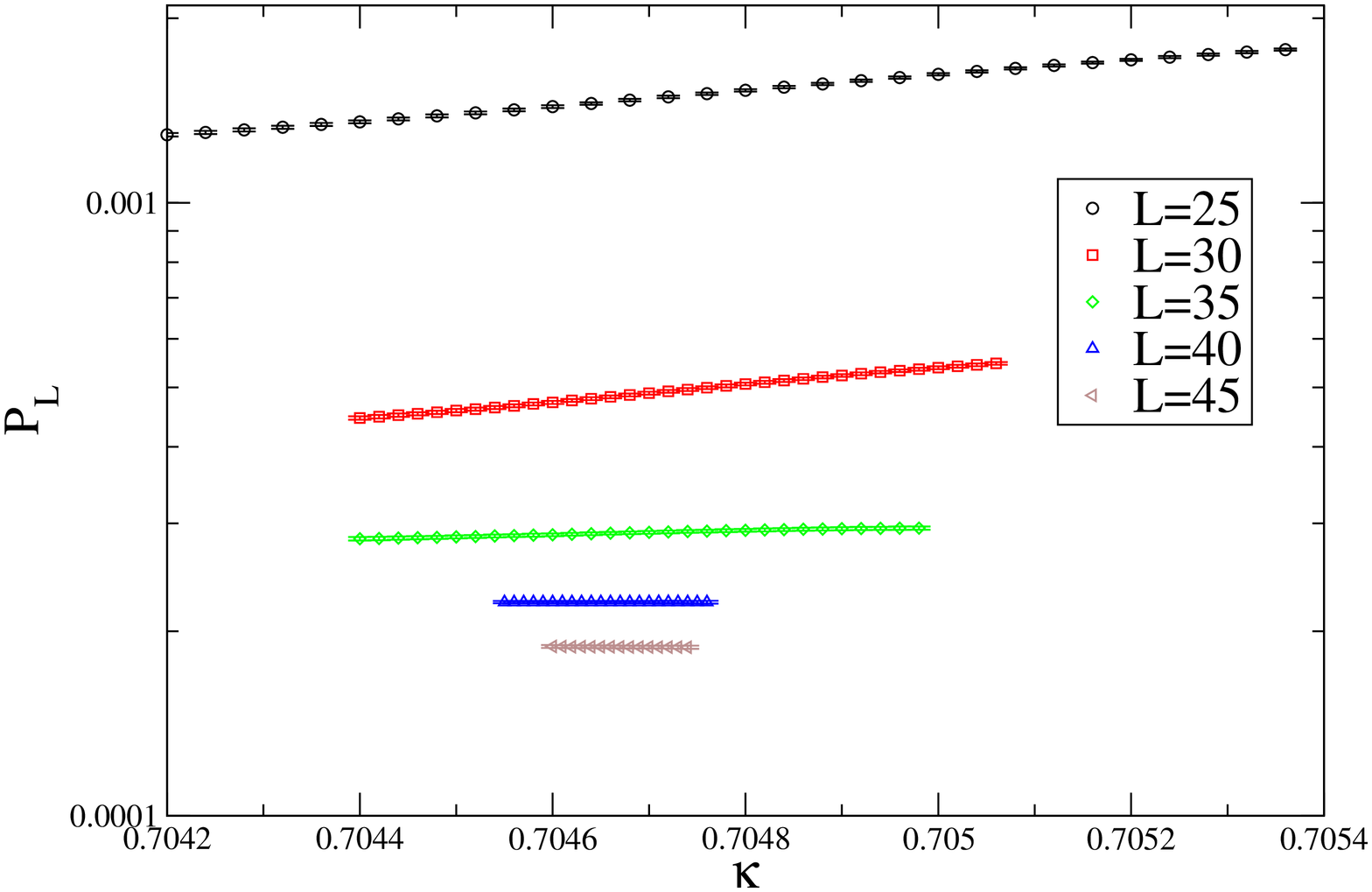}}}
\scalebox{0.29}{\rotatebox{-90}{\includegraphics{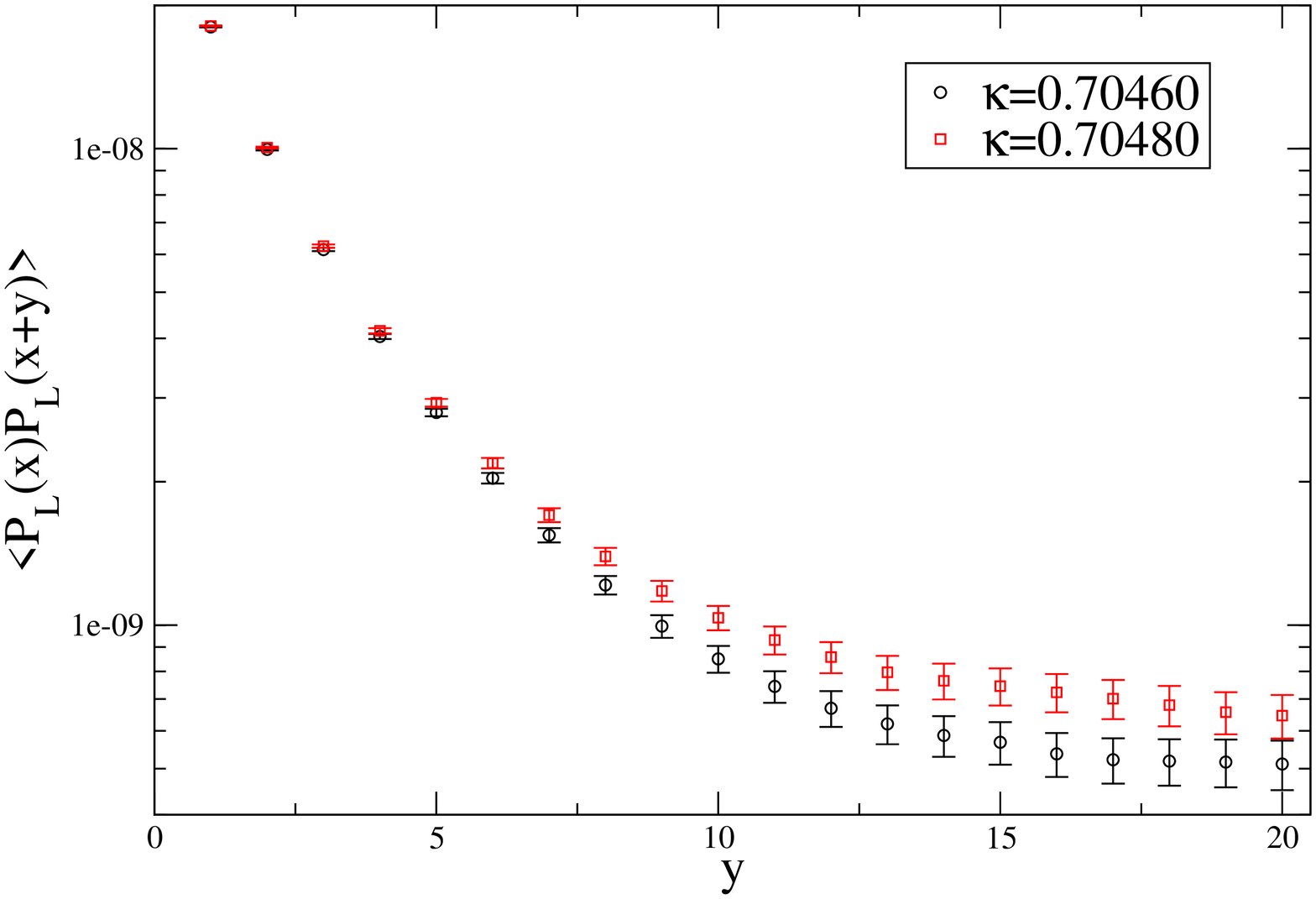}}}
\caption{\emph{Top}: Polyakov loop for \(\beta=2.775\), \(\kappa=0.70464\) and various \(L\),
the line is a fit to \(a+b \exp(-c L)\) and the result for the asymptotic value is 
\(a=1.7(1)\cdot10^{-4}\). \emph{Center)}: Polyakov loop for \(\beta=2.775\) and 
various \(L\). \emph{Bottom}: Polyakov loop correlators measured on a \(40^4\) lattice for 
\(\beta=2.775\) and \(\kappa\) slightly below (0.7046) and above (0.7048) the critical value.
 \label{fig_polyakov}}
\end{center}
\end{figure}

In this model Wilson loops never obey the area law because of the presence of the Higgs field, 
which destroys the center symmetry of the pure gauge theory; as a consequence the Polyakov loop 
is always non-zero (a direct check of this is shown in Fig.\ref{fig_polyakov}) and cannot be an 
order parameter. Nevertheless Polyakov loop is commonly used as a confinement tracker also in 
theories where it is not an order parameter, e.g. in QCD, since it abruptly jumps at the 
deconfinement transition. For the lattice \(SU(2)\) Higgs model the 
Polyakov loop doesn't seem to be influenced in any way by the transition: for 
small lattices it slightly increases, however for larger ones, the transition gets 
stronger but \(P_L\) gets flatter in the transition region, as can be seen in Fig. 
\ref{fig_polyakov} (the transition is at \(\kappa=0.704675(30)\)). Also Polyakov loop 
correlators, measured by using the multilevel algorithm of ref. \cite{LuscherWeisz}, do not show any 
significant change across the transition, as shown in the bottom of 
Fig. \ref{fig_polyakov}.

\begin{figure}[h]
\begin{center}
\setlength{\unitlength}{1cm}
\begin{picture}(5, 2.5)
\put(3,1.5){\circle*{.15}}
\put(3,1.5){\line(0,1){.5}}
\put(3,2){\line(1,0){1}}
\put(4,1.5){\line(0,1){0.5}}
\put(4,1.5){\circle*{.15}}
\put(2.2,1.6){\Large \(\langle\)}
\put(4.2,1.6){\Large \(\rangle\)}
\put(4.3,1.9){\(2\)}
\put(2.4,1.6){\(R\)}
\put(2.8,1.5){\vector(0,1){.5}}
\put(2.8,2.0){\vector(0,-1){.5}}
\put(3.3,2.3){\(2R\)}
\put(3,2.2){\vector(1,0){1}}
\put(4,2.2){\vector(-1,0){1}}
\put(2,1.2){\line(1,0){2.6}}
\put(3.1,0.0){\line(0,1){1}}
\put(3.1,1.0){\line(1,0){1}}
\put(4.1,0.0){\line(0,1){1}}
\put(3.1,0.0){\line(1,0){1}}
\put(2.3,0.4){\(2R\)}
\put(2.9,0.0){\vector(0,1){1}}
\put(2.9,1.0){\vector(0,-1){1}}
\put(3.4,-0.6){\(2R\)}
\put(3.1,-0.2){\vector(1,0){1}}
\put(4.1,-0.2){\vector(-1,0){1}}
\put(2.0,0.3){\Huge \(\langle\)}
\put(4.3,0.3){\Huge \(\rangle\)}
\put(0.2,1.1){\(O_{FM}(R)=\)}
\end{picture}
\end{center}
\caption{\(O_{FM}(R)\) definition.\label{FM_operator}}
\end{figure}

\begin{figure}[h]
\begin{center}
\scalebox{0.4}{\rotatebox{-90}{\includegraphics{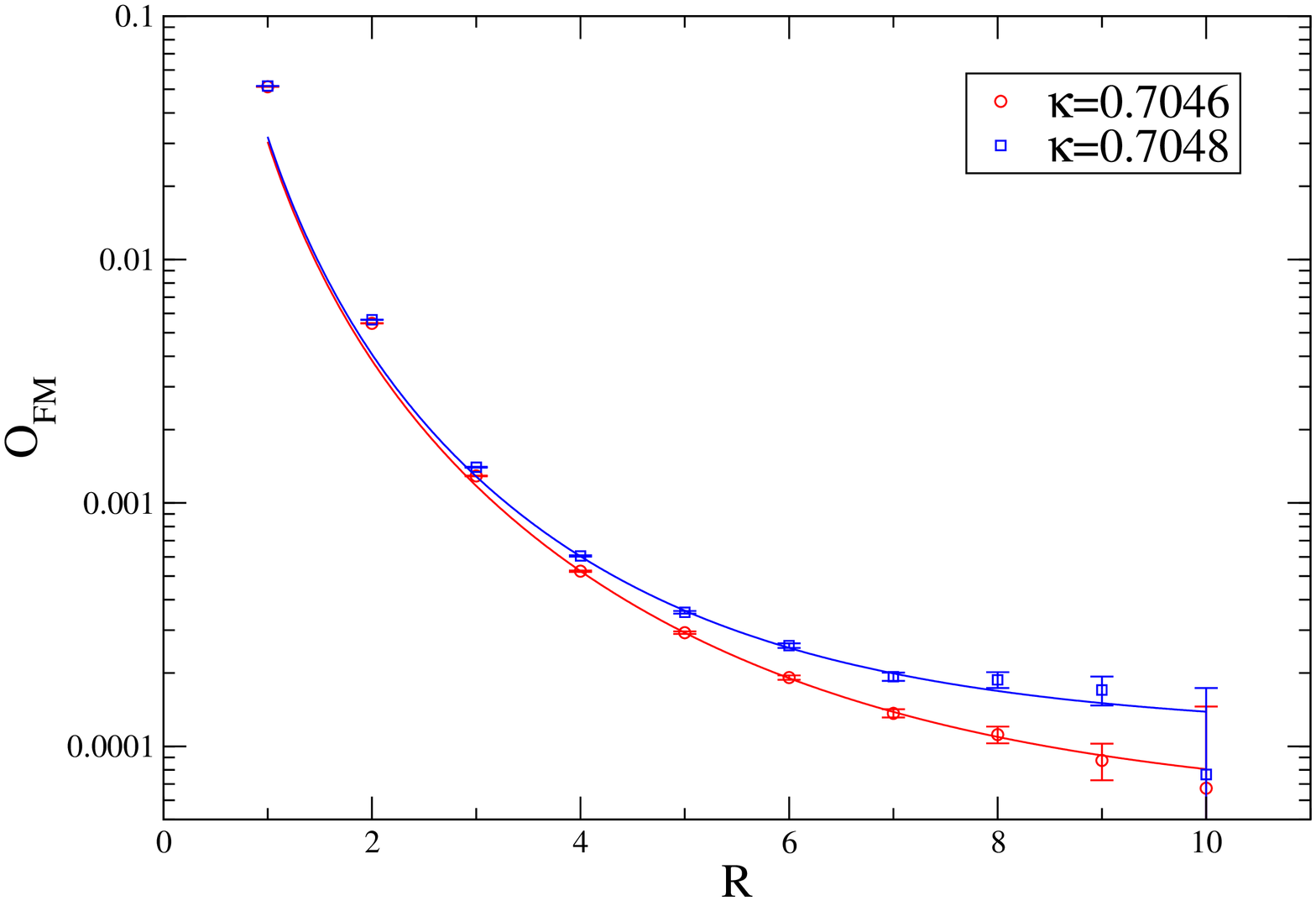}}}
\caption{Results for \(O_{FM}(R)\) at \(\beta=2.775\) on a \(40^4\) lattices slightly above and 
below the transition point.\label{fig_fm}}
\end{center}
\end{figure}

An alternative possibility is to use the order parameter introduced in \cite{FM}, which we will 
denote by \(O_{FM}\). This is defined by the limit for \(R\to\infty\) of the quantity \(O_{FM}(R)\), 
which is constructed as the square of the mean value of a staple-shaped parallel transport 
connecting two Higgs fields divided by a Wilson loop; the size of the staple is related
to that of the Wilson loop as shown in Fig. \ref{FM_operator}, where Higgs fields are represented
by dots. To our knowledge \(O_{FM}\) has never been measured in a Monte Carlo simulation, so we 
have no previous results to compare with; its strong coupling limit was computed in \cite{G2}.
\(O_{FM}(R)\) measures the overlap between a Higgs-Higgs dipole of linear dimension \(R\) and 
the vacuum; if asymptotic colored states exist they are orthogonal to the vacuum, so 
\(O_{FM}(R)\to 0\) for \(R\to \infty\), otherwise we must have \(O_{FM}(R)\to\alpha>0\). 
The results obtained by measuring \(O_{FM}(R)\) on a \(40^4\) lattice at
\(\beta=2.775\) for \(\kappa\) slightly above and slightly below the transition points are shown 
in Fig. \ref{fig_fm}, where the two lines are fits to the function\footnote{This ansatz is 
motivated by the observation that for \(R\) big enough such that the Wilson loop scales with 
perimeter, the exponentials in numerator and denominator are the same, leaving a power law.} 
\(f(x)=a+b\frac{1}{x^c}\) (symmetrized because of periodic boundary condition).

We have for the parameter \(a\), \(c\) the estimates:
\begin{eqnarray}
a_{(0.7046)}=4.9(1)\cdot 10^{-5}  & c_{(0.7046)}=2.98(5) \\
a_{(0.7048)}=10.5(5)\cdot 10^{-5} & c_{(0.7048)}=3.30(23)
\end{eqnarray}
so on both sides of the transition charge is screened and according to the interpretation of 
\cite{FM} there is color confinement.

\subsection{The pure gauge \(SU(2)\) bulk}

\begin{figure}[p]
\begin{center}
\scalebox{0.4}{\includegraphics{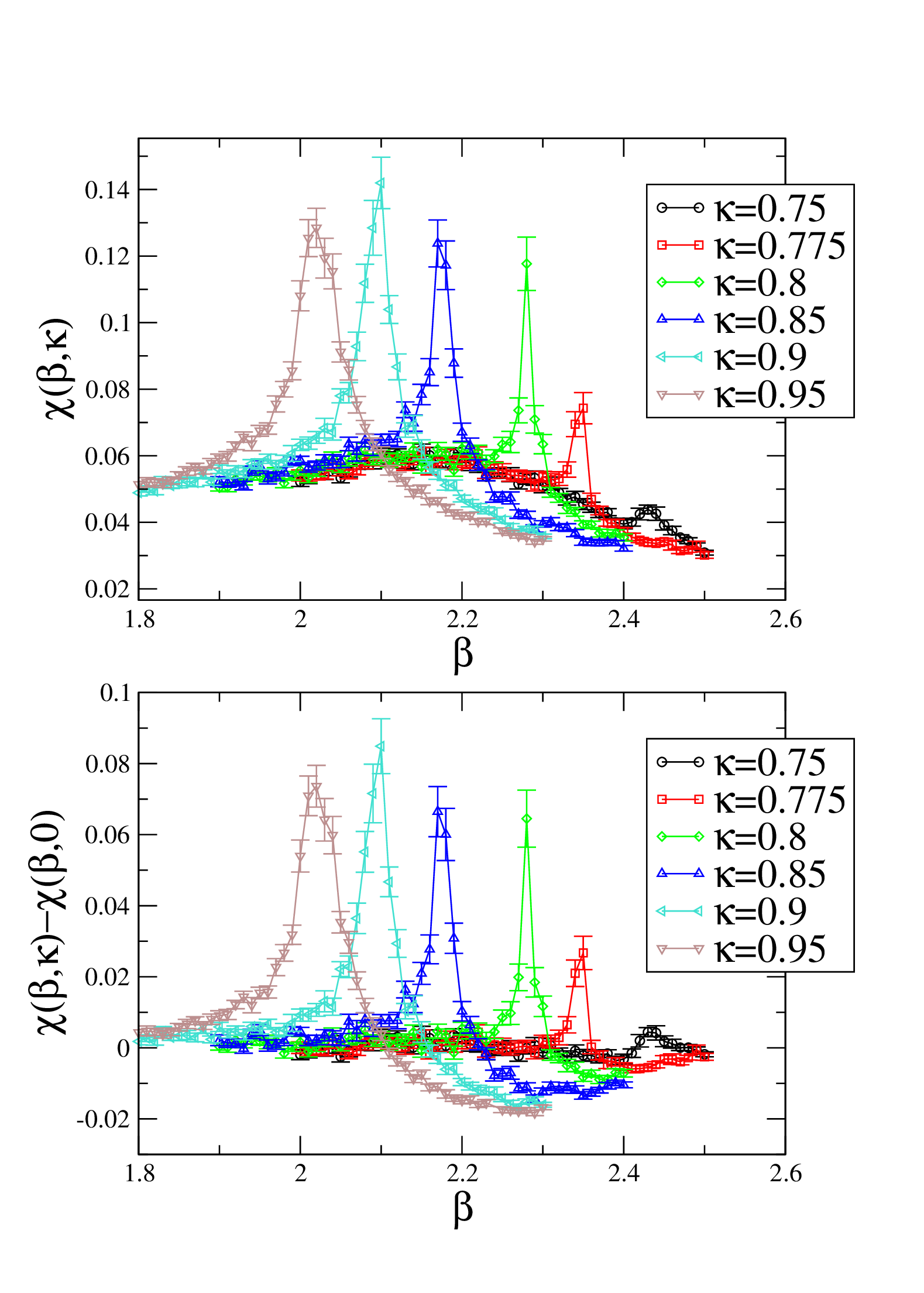}}
\caption{\emph{Up}: plaquette susceptibility. \emph{Down}: plaquette susceptibility with 
subtracted the value at \(\kappa=0\). \label{fig_plaq1}}
\end{center}
\end{figure}

Another interesting point to study is what relation (if any) the first order has with the bulk 
transition of the \(SU(2)\) pure gauge (we call this ``transition'' for the sake of simplicity, 
although it is only a rapid but smooth cross-over). An hint on the existence of such a relation
can be obtained by noting that the value \(\beta\approx 2.3\) previously thought to be the first 
order line endpoint position is the value of the bulk \(SU(2)\) transition.

Since some test simulations indicated that the position of the \(SU(2)\) bulk is very 
stable for small values of \(\kappa\) (so that in the plane \((\beta, \kappa)\) this crossover follows a 
line perpendicular to the \(\beta\) axis), it seems more convenient to hold 
\(\kappa\) fixed and vary the \(\beta\) value; the results obtained on a \(30^4\) lattice for the 
plaquette susceptibilities for several \(\kappa\) values are shown in Fig. \ref{fig_plaq1}.

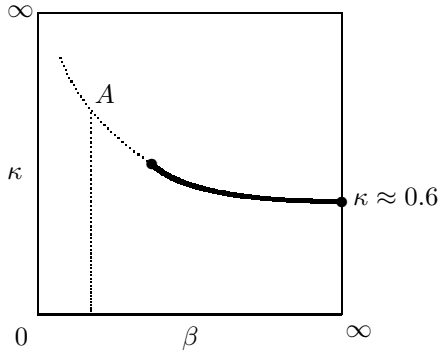
\begin{figure}[h]
\begin{center}
\setlength{\unitlength}{1cm}
\begin{picture}(4, 4)
 \put(0, 0){\line(0, 1){4}}
 \put(0, 0){\line(1, 0){4}}
 \put(0, 4){\line(1, 0){4}}
 \put(4, 0){\line(0, 1){4}}
 \put(4, 1.5){\circle*{.15}}
 \put(1.5, 2){\circle*{.15}}
 \put(1.9,-.4){\(\beta\)}
 \put(-.3,-.4){\(0\)}
 \put(4.05,-.3){\(\infty\)}
 \put(-.4,1.8){\(\kappa\)}
 \put(-.4,3.95){\(\infty\)}
 \qbezier[30](0.3,3.4)(0.6,2.6)(1.5,2)
 \qbezier[50](0.7,0)(0.7,1)(0.7,2.675)
 \put(0.75,2.8){\(A\)}
 \put(4.15, 1.45){\(\kappa\approx 0.6\)}
 \linethickness{0.5mm}
 \qbezier(1.5, 2)(2, 1.5)(4, 1.5)
\end{picture}
\end{center}
\caption{Phase diagram of the Higgs \(SU(2)\) with rapid crossovers indicated by dotted lines.
\label{fig_cross}}
\end{figure}

\begin{figure}[h]
\begin{center}
\scalebox{0.4}{\rotatebox{-90}{\includegraphics{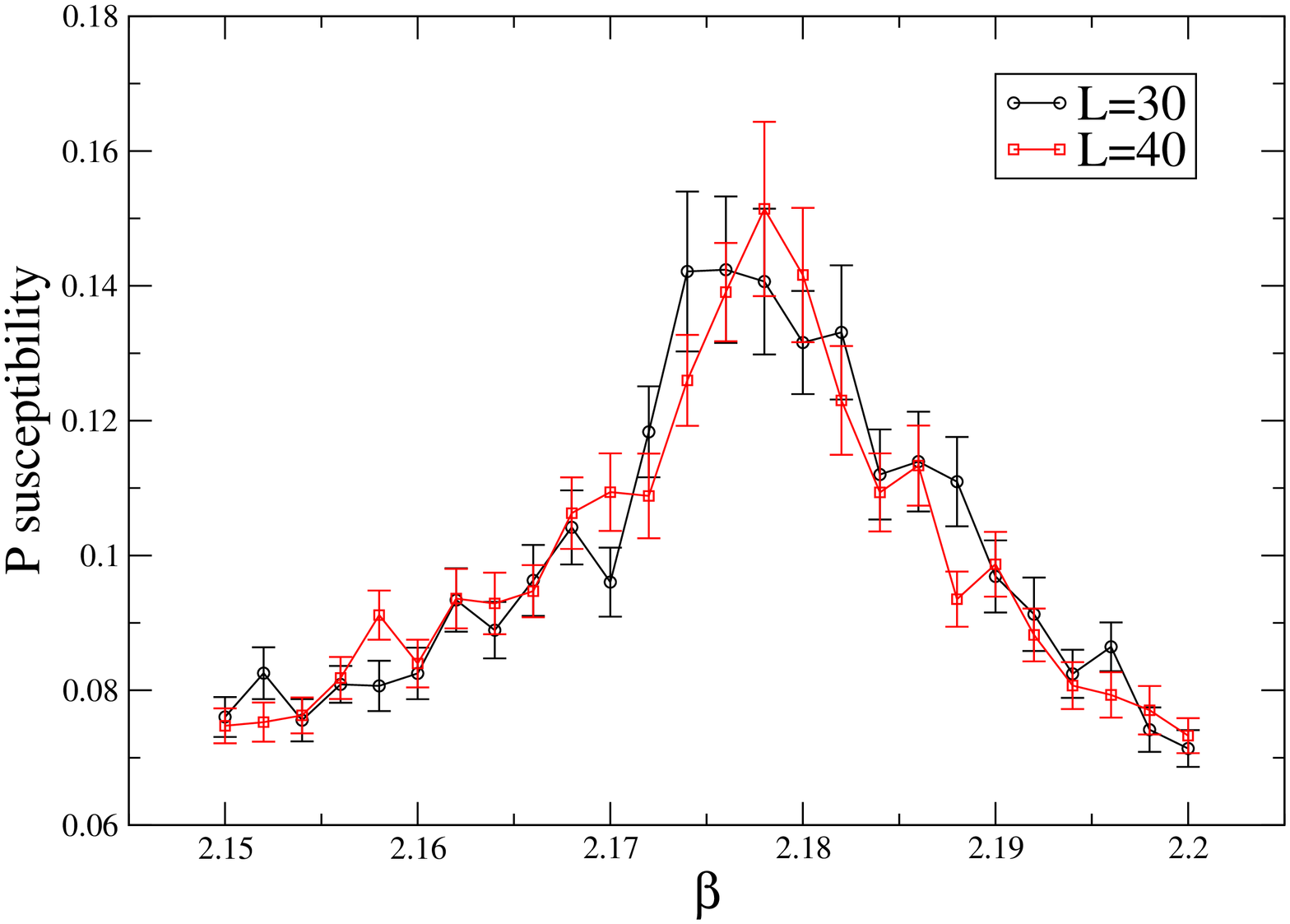}}}
\caption{Plaquette susceptibility for \(\kappa=0.85\). \label{fig_plaq2}}
\end{center}
\end{figure}

From these data a structure of cross-over lines emerges as shown in Fig. \ref{fig_cross}. 
For \(\kappa<0.6\) the bump in \(\beta\) of the plaquette susceptibility which corresponds to
the \(SU(2)\) bulk transition is independent on \(\kappa\) both in position and in shape
(vertical dotted line in Fig. \ref{fig_cross}).
 For \(\kappa\) larger than \(0.6\)  two peaks in \(\beta\) appear, the bulk one and the 
first order transition studied above. The latter peak persists also when \(\kappa\) is greater 
than the critical value corresponding to the endpoint andthere is no real transition, 
although it is smaller that the \(SU(2)\) bulk signal (see Fig. \ref{fig_plaq1}, 
\(\kappa=0.75\)). By increasing the \(\kappa\) value the cross-over remnant of the first 
order transition moves towards smaller \(\beta\) values, until it intersects the 
\(SU(2)\) bulk in the point indicated by \(A\) in Fig. \ref{fig_cross}. In the neighborhood
of the point \(A\) the ``first order continuation'' peak gets stronger and the \(SU(2)\) bulk
disappears. This increase could have been misinterpreted as the first order transition line
endpoint in previous works. A direct check to ensure that in this region there is no transition 
is shown in Fig. \ref{fig_plaq2}.  For still larger \(\kappa\) values only one maximum is 
present in susceptibilities, which gets weaker as \(\kappa\to \infty\). 

This interplay between the first order transition line and the bulk \(SU(2)\) suggests 
that the first order transition could be in some way related to the same lattice artifacts that 
drive the bulk \(SU(2)\) transition, namely the \(Z_2\) monopoles. The density of \(Z_2\) 
monopoles seems indeed to have a jump across the transition (see Fig. \ref{fig_max2.775}).

\subsection{Finite temperature}

Motivated by the last observation we try to investigate if the first order transition line 
can itself be thought as a lattice artefact, like the pure gauge \(SO(3)\) first order bulk 
transition. To answer this question the simplest method is to consider the system at finite 
temperature: bulk transitions are insensitive to the temporal extension \(L_t\) of the lattice, 
while physical transitions scale by varying \(L_t\). 

\begin{figure}
\begin{center}
\scalebox{0.4}{\rotatebox{-90}{\includegraphics{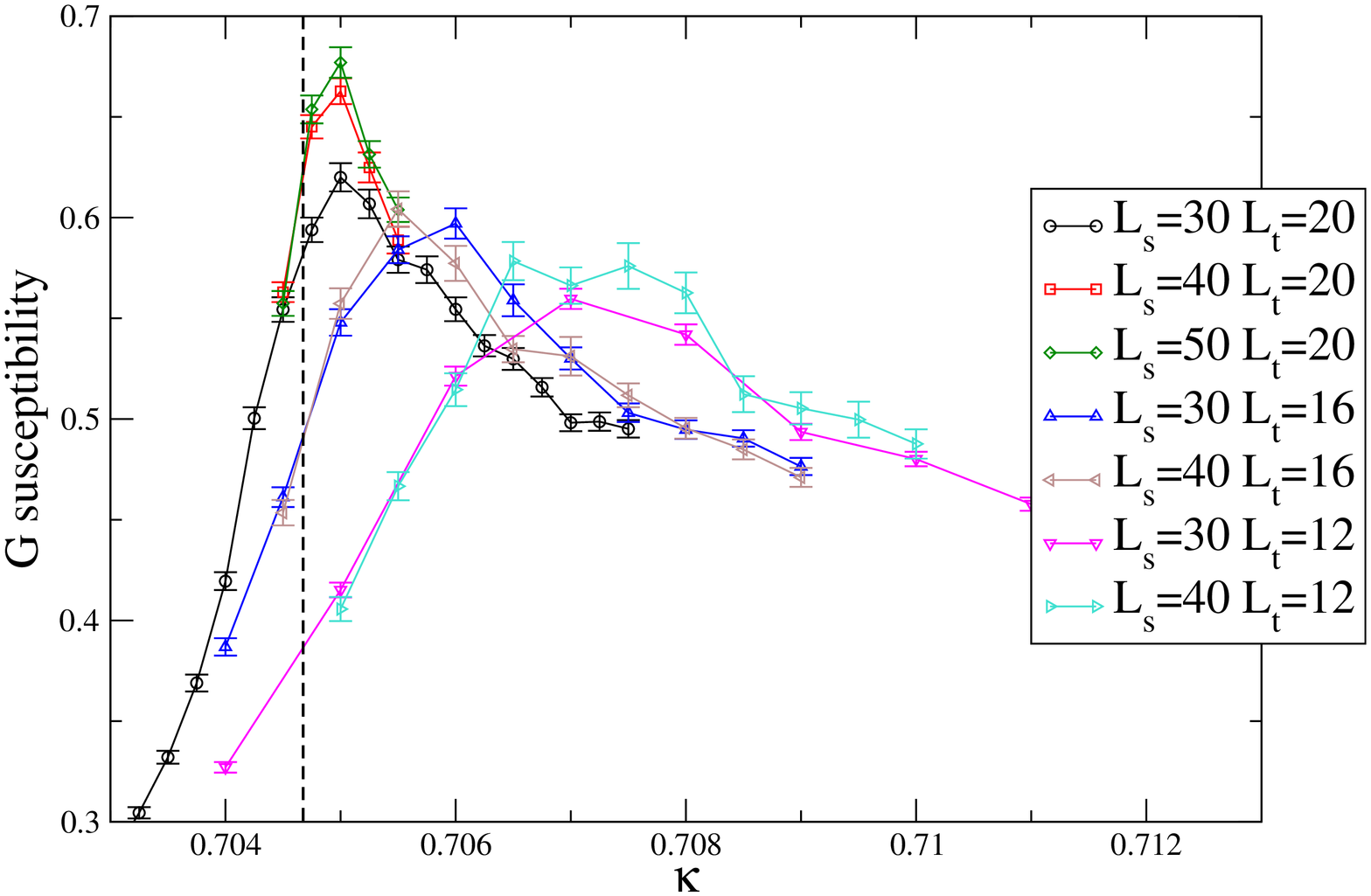}}}
\caption{G susceptibility for \(\beta=2.775\); the vertical line indicates the place where on 
symmetric lattices the transition is observed.\label{fig_finitet}}
\end{center}
\end{figure}

For each \(L_t\) value we simulate the system with several spatial extents \(L_s\), in order to 
perform a finite size analysis. Some of the results obtained are shown in Fig. \ref{fig_finitet}.
It is clear that the position of the maxima of the \(G\) susceptibility moves by varying 
\(L_t\), so that the line of first order transition seen at zero temperature cannot be a 
lattice artifact. Moreover for all the values of \(L_t\) considered the maxima saturate 
by increasing \(L_s\), indicating the absence of a phase transition. We expect that for 
sufficiently small temperature (i.e. large enough \(L_t\)) the first order transition is restored, 
however to observe this behaviour numerically we should use lattices with 
\(L_t>20\) and \(L_s \gg L_t\), typically \(L_s\gtrsim 3 L_t\). Instead we have to 
restrict ourselves to  \(L_s^3L_t \lesssim 4\times 10^6\) because of computer capability. 
In any case we see the cross-over gets stronger by increasing the lattice temporal extension.

\section{Conclusions}
This paper is a study of the phase diagram of the \(SU(2)\) Higgs gauge 
theory with the Higgs in the fundamental representation and fixed length. We find that to investigate 
the phase structure of this system it is necessary to use much larger lattices than the ones adopted in 
the past due to the large correlation length.

We present the first clear evidence of the presence of a line of first order transition and we estimate 
its endpoint position to be at \(\beta_c=2.7266(16)\), thus confirming an expectation based on the 
Fradkin-Shenker theorem. 

We give indications that this transition is not a deconfinement transition. 

We attempt an exploration of the system at finite temperature; the first order transition becames then 
a cross-over whose position moves by varying the temperature.

\section{Acknowledgments} 
Simulation were performed using the Italian GRID Infrastructure and one of us (C.~B.) wishes to 
thank Tommaso Boccali for his technical assistance in its use. We thank A.~D'Alessandro for 
substantial help in the early stages of this work.

\appendix

\section{Data}

In this appendix we give some details of our numerical results. The notation is as follows
\begin{itemize}
\item \(\kappa_{pc}\) is the location in \(\kappa\) of the \(G\) susceptibility maximum
\item \(\chi_O\) is the peak value of the susceptibility of the \(O\) observable 
\item \(B\) is the value at minimum of the energy Binder cumulant 
\end{itemize}
        
\begin{center}
\begin{tabular}{|c|l|l|l|l|l|l|}
\hline
\multicolumn{7}{|c|}{\(\beta=2.725\)}\\      
\hline
\(L\) & \(\kappa_{pc}\) & \(\chi_G\) & \(\chi_P\) & \(\chi_M\) & \(\chi_E\) & 2/3-\(B\) \\
\hline  
20 & 0.7090(2) & 0.615(18) & 0.005575(88)  & 0.1147(19) & 10.85(30) & 6.36(18)e-07 \\
25 & 0.7089(2) & 0.692(16) & 0.005831(71)  & 0.1194(15) & 11.89(26) & 2.839(62)e-07 \\
30 & 0.7088(1) & 0.858(20) & 0.006485(78)  & 0.1319(15) & 14.51(32) & 1.679(38)e-07 \\
35 & 0.70873(5) & 1.018(24) & 0.007082(89)  & 0.1431(17) & 16.64(37) & 1.039(23)e-07 \\
40 & 0.70866(5) & 1.022(24) & 0.00704(12)   & 0.1426(18) & 16.85(37) & 6.17(13)e-08 \\
45 & 0.70886(5) & 0.994(31) & 0.00725(23)   & 0.1469(25) & 16.25(47) & 3.71(10)e-08 \\
\hline
\end{tabular}
\end{center}

\begin{center}
\begin{tabular}{|c|l|l|l|l|l|l|}
\hline
\multicolumn{7}{|c|}{\(\beta=2.775\)}\\      
\hline
\(L\) & \(\kappa_{pc}\) & \(\chi_G\) & \(\chi_P\) & \(\chi_M\) & \(\chi_E\) & 2/3-\(B\) \\
\hline  
25 & 0.7049(1)  & 0.5740(65) & 0.005459(31) & 0.10807(62) & 9.460(98) &  2.156(22)e-07 \\
27 & 0.70480(8) & 0.6211(82) & 0.005528(40) & 0.10884(74) & 10.16(12) &  1.702(21)e-07 \\
30 & 0.70476(5) & 0.734(13)  & 0.005769(52) & 0.1135(10)  & 11.82(19) &  1.299(21)e-07\\
35 & 0.70466(5) & 0.812(16)  & 0.006044(64) & 0.1182(12)  & 12.88(24) &  7.64(14)e-08\\
40 & 0.70469(3) & 0.956(17)  & 0.006573(77) & 0.1285(12)  & 14.66(25) &  5.099(88)e-08\\
45 & 0.704675(30) & 1.114(24)& 0.00698(10)  & 0.1359(16)  & 16.72(34) &  3.631(75)e-08\\
\hline
\end{tabular}
\end{center}

\begin{center}
\begin{tabular}{|c|l|l|l|l|l|l|}
\hline
\multicolumn{7}{|c|}{\(\beta=2.79\)}\\      
\hline
\(L\) & \(\kappa_{pc}\) & \(\chi_G\) & \(\chi_P\) & \(\chi_M\) & \(\chi_E\) & 2/3-\(B\) \\
\hline  
25 & 0.70380(10) & 0.5512(97) & 0.005301(48) & 0.10362(93) & 9.05(14)  & 2.030(32)e-07 \\
30 & 0.70360(5) & 0.646(12)  & 0.005481(49) & 0.10632(95) & 10.39(17) & 1.125(19)e-07 \\
40 & 0.70356(3) & 0.912(19)  & 0.006187(78) & 0.1200(13)  & 14.01(27) & 4.799(93)e-08 \\
45 & 0.703525(30) & 1.140(23)  & 0.006901(97) & 0.1318(15)  & 17.01(33) & 3.639(70)e-08 \\
\hline
\end{tabular}
\end{center}

\begin{center}
\begin{tabular}{|c|l|l|l|l|l|l|}
\hline
\multicolumn{7}{|c|}{\(\beta=2.8\)}\\      
\hline
\(L\) & \(\kappa_{pc}\) & \(\chi_G\) & \(\chi_P\) & \(\chi_M\) & \(\chi_E\) & 2/3-\(B\) \\
\hline  
25 & 0.7032(2) & 0.683(22)  & 0.005080(80) & 0.0978(15) & 11.25(34) & 2.500(75)e-07 \\    
30 & 0.70298(8) & 0.7212(85) & 0.005279(38) & 0.10181(60)& 11.63(13) & 1.246(13)e-07 \\
35 & 0.70286(5) & 0.760(16)  & 0.005685(62) & 0.1093(11) & 11.64(22) & 6.74(13)e-08 \\ 
40 & 0.70282(4) & 0.946(22)  & 0.006239(75) & 0.1192(13) & 14.36(31) & 4.87(11)e-08 \\
45 & 0.70279(3) & 1.219(27)  & 0.007033(89) & 0.1339(16) & 17.75(37) & 3.759(78)e-08 \\
\hline
\end{tabular}
\end{center}

\end{document}